%% file: conference_101719.tex
\documentclass[conference]{IEEEtran}
\IEEEoverridecommandlockouts
\usepackage{cite}
\usepackage{amsmath,amssymb,amsfonts}
\usepackage{algorithmic}
\usepackage{graphicx}
\usepackage{textcomp}
\usepackage{xcolor}
\usepackage{comment}
\usepackage{url}

\def\BibTeX{{\rm B\kern-.05em{\sc i\kern-.025em b}\kern-.08em
    T\kern-.1667em\lower.7ex\hbox{E}\kern-.125emX}}
\begin{document}

\title{A next-generation platform for Cyber Range-as-a-Service}

\author{\IEEEauthorblockN{Vittorio Orbinato}
\IEEEauthorblockA{\textit{DIETI, Università degli Studi di Napoli Federico II}, Naples, Italy \\
vittorio.orbinato@unina.it}}

\maketitle

\input{tex/abstract}


\input{tex/introduction}

\input{tex/background}

\input{tex/proposal}

\input{tex/conclusion}

\input{tex/acknowledgments}

\bibliographystyle{plain}
\bibliography{biblio}

\end{document}

%% file: tex/abstract.tex
\begin{abstract}
In the last years, Cyber Ranges have become a widespread solution to train professionals for responding to cyber threats and attacks. Cloud computing plays a key role in this context since it enables the creation of virtual infrastructures on which Cyber Ranges are based. However, the setup and management of Cyber Ranges are expensive and time-consuming activities. In this paper, we highlight the novel features for the next-generation Cyber Range platforms. In particular, these features include the creation of a virtual clone for an actual corporate infrastructure, relieving the security managers from the setup of the training scenarios and sessions, the automatic monitoring of the participants' activities, and the emulation of their behavior.
\end{abstract}

%% file: tex/introduction.tex
\section{Introduction}
\label{sec:introduction}

NIST defines Cyber Ranges as ``\textit{interactive, simulated platforms and representations of networks, systems, tools, and applications that are connected to a simulated Internet level environment}". They provide a safe, legal environment to gain hands-on cyber skills and a secure environment for product development and security posture testing. A Cyber Range may include actual hardware and software or be a combination of actual and virtual components" \cite{NIST-CR-Guide}.
Nowadays, there are several technologies capable of supporting Cyber Ranges. The majority of them are based on cloud computing to create virtual infrastructures, including hosts, subnets, and services, which define the training environment. Cloud computing enables the use of large data centers at affordable costs to simulate large networks. Examples of these solutions are KYPO \cite{KYPO}, the National Cyber Range \cite{NCR}, Redcloud \cite{Redcloud}.

However, the creation and the management of Cyber Ranges are cumbersome tasks, in terms of time and costs. Indeed, Cyber Range organizers need to create and configure the machines and networks to be used for training, to deploy services, data, and workloads that emulate the actual infrastructure of the organization realistically. Moreover, they have to deal with the analysis of the participants' activities to provide them feedback, and engage large groups of people from the company to take the roles required by the Cyber Range (e.g., blue teams for defense, red teams for the attack, white teams for monitoring). 
All these activities involve a significant amount of personnel (e.g., in terms of man-hours), which makes the world of the Cyber Ranges not easily approachable for companies and organizations.

We strongly believe that the next generation Cyber Ranges should reduce the effort of the managers in the setup of these platforms and ease the management and monitoring of all the related activities to work in a better way. Hence the need to introduce new technologies to overcome the limitations to the adoption and diffusion of Cyber Ranges.
This paper focuses on the next generation Cyber Range platform, which should provide the creation and management of Cyber Ranges in an efficient way and with limited costs. The platform will be based on: 
\begin{itemize}
    \item Advanced hypervisor-level technologies for monitoring and analyzing the activities of the participants; 
    \item Statistical and AI techniques to learn the features of corporate infrastructure and automatically generate scenarios based on that information;
    \item AI techniques to simulate the participants to allow smaller teams to use Cyber Ranges.
\end{itemize}

The solution will enable exercise environments with high complexity (in terms of number and heterogeneity of services, network elements, etc.) to emulate the ICT systems realistically employed by companies. 
These emulated infrastructures are meant to be highly flexible, easy to manage and configure (e.g., by providing visual interfaces), and to allow managers to monitor the activities of the users. 

In the reminder of this paper, Section~\ref{sec:Background} shows a background analysis of Cyber Ranges along with the open research problems, Section~\ref{sec:Proposal} discusses the research proposal, Section~\ref{sec:Conclusion} concludes the paper.

%% file: tex/background.tex
\section{Background}
\label{sec:Background}

Cyber Ranges can be classified into different categories, based on their features and capabilities. NIST identifies four main types of Cyber Ranges in its Cyber Range Guide \cite{NIST-CR-Guide}: 

\begin{itemize}
    \item \textbf{\textit{Simulation Ranges}} run in virtual instances, without the need for any physical network equipment. Virtual machines (VM) are used to replicate the server, network, and storage configuration of specific infrastructure. These VMs are based on standardized templates, so their fidelity in replicating the target infrastructure may be limited. 
    \item \textbf{\textit{Overlay Ranges}} run on top of real networks. They grant a more significant level of fidelity than Simulation Ranges but, in terms of hardware, the cost is greater, along with the risk of compromising the underlying network infrastructure. Overlay Ranges grant very low flexibility, because a slight change in the configuration may cause the redefinition of the entire Cyber Range. An example of an overlay network is the Global Environment for Network Innovations (GENI) \cite{GENI}, sponsored by the National Science Foundation. 
    \item \textbf{\textit{Emulation Ranges}} run on dedicated network infrastructures, mapping the specific network/server/storage configuration onto them. The physical infrastructure becomes the Cyber Range itself. Emulation Ranges include traffic generation, with the possibility of emulating traffic flows, specific patterns, attacks, etc. The National Cyber Range (NCR) \cite{NCR}, the Michigan Cyber Range \cite{MichiganCR} and DETER \cite{DETER} are examples of the Emulation Range initiative. 
    \item \textbf{\textit{Hybrid Ranges}} can be defined as a customized combination of the aforementioned types of Cyber Ranges: they combine the advantages of the other types of Ranges, such as the high fidelity of Overlay Ranges and the flexibility of Emulation Ranges. An example of this type of Range is the Virginia Cyber Range \cite{VirginiaCR}.
\end{itemize}

Yamin \textit{et al.} \cite{AYAMIN2020101636} proposed a different classification of the Cyber Ranges, defining a taxonomy of the main Cyber Range-related concepts: scenarios, monitoring, and management.

\vspace{0.2cm}
\noindent
\textit{\textbf{Scenario}}. It defines the execution context of a training exercise. and can be seen as the combination of the following elements: purpose, environment, storyline, type, and domain. 
The \textit{purpose} explains the goals of the scenario, which are necessary to develop the environment. The \textit{environment} is the topology in which the scenario is executed, it can be either a technical infrastructure or a non-computer-based one. 
The \textit{storyline} tells one or more stories about how the exercise is executed and describes the events and actions that constitute the scenario and how these are linked to define the narrative. The \textit{type} of the scenario influences the nature of the environment and the storyline. The type of a scenario can be static or dynamic. 
Also, the \textit{domain} of the scenario is an essential aspect: some examples of application domain are IoT, network, cloud, etc.

\vspace{0.2cm}
\noindent
\textit{\textbf{Monitoring}}. It is the set of methods, tools, and layers at which real-time monitoring of training sessions is performed \cite{staff2012joint}.  The methods employed to monitor training exercises can be automatic, thanks to the use of specific tools (intrusion detection systems, security information, and event management, etc.), or manual, performed by human observers. Monitoring can be performed at different layers. Depending on the type of exercise, it is possible to monitor the session at multiple TCP/IP layers or an abstract social layer.

\vspace{0.2cm}
\noindent
\textit{\textbf{Management}}. The management of a Cyber Range involves many different activities, from role management during training sessions to resource and range management. All of these aspects of management impact on the development of a training exercise: role management is related to the assignment of roles to individuals and teams during a training exercise, resource management deals with the allocation of memory, disk, and network resources needed to perform the exercise and, finally, range management is associated to how the results of the training exercise are presented to the managers. 
In a Cyber Range environment, \textit{\textbf{teaming}} represents the group of professionals that design, develop, manage and participate in a training session. Teams are identified by different colors according to their roles during the exercise.
The \textit{red team} is the one that plays the role of the attackers, trying to violate the security of the environment by exploiting its vulnerabilities. 
The \textit{blue team} covers the defensive role, with the responsibility to identify and fix potential vulnerabilities that can be exploited by the red team.
The \textit{white team} designs the scenario organizes the exercise and sets the rules of engagement between the red and blue team.

%% file: tex/proposal.tex
\section{Proposal}
\label{sec:Proposal}

The research activity presented in this paper aims to develop new Cyber Range techniques which will be integrated into a next-generation platform. This platform will have the following properties:
\begin{itemize}
    \item Instantiation of large scale virtual network scenarios, using several amounts of computing resources available in the datacenters through cloud computing;
    \item Clonation of a corporate infrastructure inside the virtual scenario, to simulate realistically the users, resources, processes, and network traffic of such infrastructure. This replica will be used for training exercises without impacting on the actual one;
    \item Ease of management for new training scenarios (e.g., new network topologies, vulnerabilities configuration, etc.) by providing the managers with a high-level domain-specific language to describe and design highly complex scenarios, along with APIs and visual dashboards;
    \item Silent and automatic monitoring of the activities of the participants, using Virtual Machine Introspection (VMI) techniques at the hypervisor level, in order to provide the managers and the participants with detailed feedback;
    \item Actor simulation (attackers/defenders) through AI techniques, in order to help the staff involved in the training and allow small groups of people to complete training sessions successfully, even when they cannot cover all the required roles.
\end{itemize}

\begin{figure}[htbp]
    \centering
    \includegraphics[width=1.0\linewidth]{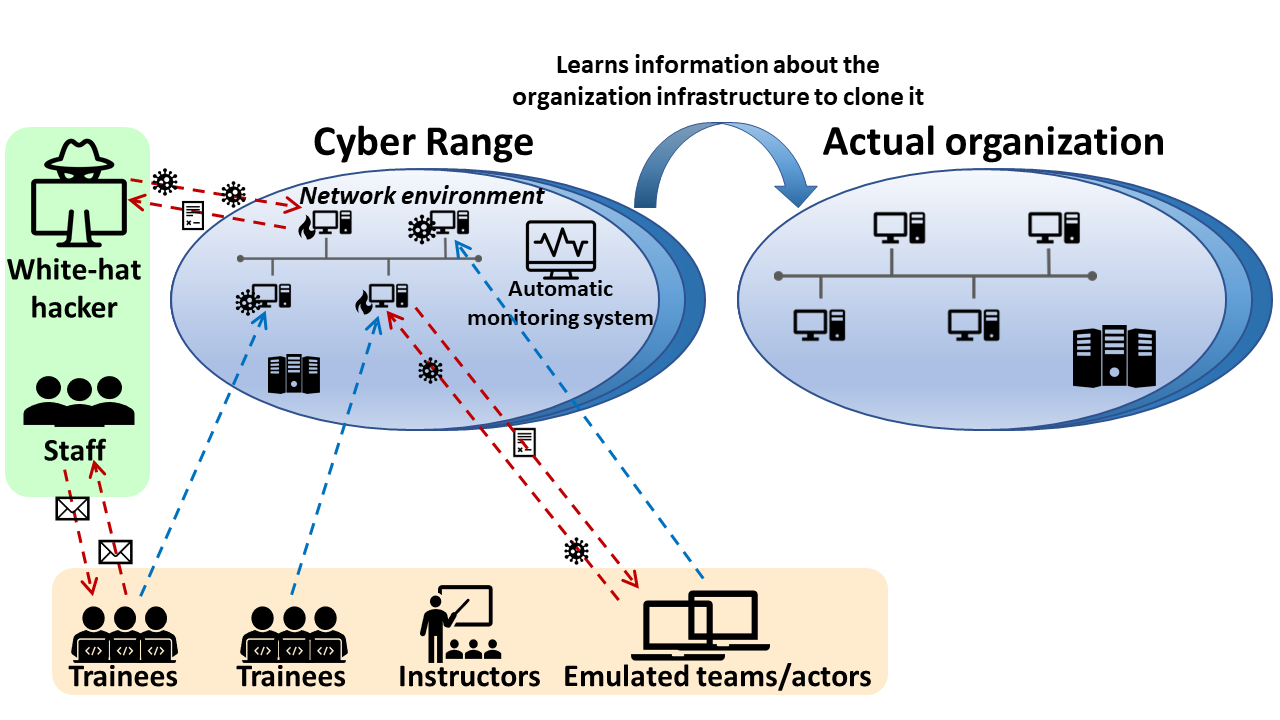}
    \caption{Overview of the next-generation Cyber Range platform}
    \label{fig:Cyber Range infrastructure}
\end{figure}

\figurename{}~\ref{fig:Cyber Range infrastructure} shows an overview of the next generation Cyber Range platform. It is possible to notice how the Cyber Range replicates the actual infrastructure, after learning its features, providing a virtual clone of it. In this way, all the training exercises will be performed without impacting the actual infrastructure. The replica also provides an automatic monitoring system, in order to silently track all the events that occur in the training sessions and the activities of the participants. Moreover, the Cyber Range offers the possibility to support the staff (i.e., instructors and trainees) involved in the training sessions with emulated teams and actors.

The innovation degree of our proposal can be considered extremely high: from the technological point of view because of the development of new systems based on cloud computing and hypervisor technologies; from the methodological one because these solutions will leverage statistical and AI techniques to improve the realism of the training exercises.
The research activity will be carried out in three different phases, explained in the following.

\subsection{Advanced scenario simulation}
\label{sec:phase1}

The first phase of the research activity will focus on the replication of corporate infrastructure, creating a \textit{digital twin} \cite{systems7010007}, \cite{8477101}. This replica will be easy to access and use, in terms of management, configuration, and monitoring, according to the "as-a-service" paradigm. 
Creating a digital twin of a corporate infrastructure means defining and building a virtual infrastructure that simulates realistically its users, resources, processes, and network traffic. This virtual infrastructure will be defined using a statistical model, which will capture information on the services features and network topologies in order to replicate similar features in the virtual infrastructure. The digital twin will be created by instantiating large-scale virtual scenarios, using huge amounts of computing resources supplied by the companies data centers through cloud computing technologies. The emulation of the infrastructure and its users, resources, processes, and network traffic should be as realistic as possible, with the possibility to perform reliable security training without impacting on the real one. 

The main goal of this phase is the realization of a custom engine, which allows the managers to model complex training scenarios using a domain-specific language. This high-level language will be used to design the services, resources, behaviors, and vulnerabilities. The engine will convert the high-level description of the scenarios in low-level operations on the infrastructure, through the state-of-the-art technologies for Infrastructure-as-a-Service and Infrastructure-as-code. It will be based on the cloud computing platform OpenStack \cite{OpenStack}, a de facto standard technology, using its API to create complex network scenarios (hundreds of network elements, e.g., hosts, routers, switches, firewalls, etc.). 
Moreover, the engine will be equipped with a repository of host configurations to emulate network servers and clients that will generate realistic traffic flows. To achieve a high level of realism, the platform will support the profiling of the existing corporate infrastructures to create a statistical descriptive model, which will be used to set up a virtual infrastructure. In this way, the Cyber Range managers will be facilitated in designing realistic training scenarios, without the need of modeling and replicating every single detail of the target infrastructure.

The statistical model will grant the confidentiality of sensible data, because the simulation will generate a different and anonymized version of the infrastructure, along with its users, resources, and traffic. 
This task will be accomplished with the integration of advanced techniques to simulate realistically the users, resources, processes, and network traffic of the corporate infrastructure. These techniques will be based on the data gathered from the target infrastructure, which will be used to define a statistical model that reflects the behavior of the users and the resource distribution. Both statistical techniques, such as graphs and Markovian models, and modern techniques, like deep neural networks, will be taken into account to complete this task. The model will be used to generate training scenarios on the virtual infrastructure, which will be representative of the actual one but also anonymized to protect sensitive information. The simulated scenarios will be easy to manage, thanks to visual dashboards and APIs, fit to support a Cyber Range-oriented workflow. The dashboards will allow the managers to visually interact with the virtual infrastructures, providing them with a set of configurable vulnerabilities which can be used to create new scenarios in which the red and blue teams will challenge each other.

\subsection{Monitoring techniques}
\label{sec:phase2}
The aim of the second phase is the creation of a hypervisor-based monitoring system to track and analyze the activities of the red and blue teams. These activities, such as connection to a specific node, attack traffic generation, malicious code execution, etc. will be monitored silently, through Virtual Machine Introspection techniques. VMI describes the act of examining and monitoring the virtual machine (i.e., the state of the guest
OS) from the vantage point of the hypervisor or a second privileged
guest OS \cite{pfoh2009formal}. Three properties of virtualization enable and support Virtual Machine Introspection, namely isolation, inspection, and interposition.
Specifically, isolation is the property that states that ideally the hypervisor cannot be tampered with; inspection refers to the property that allows the VMM to examine the entire state of the guest OS without having to rely on a possibly compromised system for this information; and, finally, interposition is the
ability to inject operations into the normal running of the system based on certain conditions \cite{garfinkel2003virtual}. To be able to rely on the advantages that VMI brings, the hypervisor must not be vulnerable to compromise. Due to the strong isolation properties provided by the hypervisor, the number of attack vectors on the external monitoring components is significantly reduced compared to software running inside the monitored VM \cite{pfoh2009formal}. 

The hypervisor enables the interception of events in the guest OS, by taking advantage of dynamic probing mechanisms provided by most modern commodity OSes (such as DTrace for FreeBSD, Mac OS X, and Solaris \cite{gregg2011dtrace},  SystemTap and Kprobes for Linux \cite{cohen2005instrumenting}, \cite{cohen2005gaining}, and Detours for Microsoft Windows \cite{Detours}) \cite{8390705}. These mechanisms can collect information from the kernel through the use of specific functions and breakpoints: for example, a breakpoint can be used to probe the
invocations of an API function. The run-time
the overhead of these probing mechanisms is low enough to be applicable in many applications: for example, benchmarks
from VMware reported that, in the worst cases, dynamic probes
cause a throughput loss between 3\% and 4.5\% \cite{carbone2014vprobes}.

These techniques will make monitoring activities non-intrusive and invisible to the teams, in order to grant a higher level of realism during training. Moreover, these techniques will relieve the managers of the host management burden, allowing them to avoid the installation of new processes or the alteration of the network configuration. The monitoring system will be implemented as modules for the Linux/KVM hypervisor \cite{KVM}, or other hypervisors of interest, e.g., VMware ESXi \cite{VMware-ESXi}.
The data gathered by the monitoring system will be used to automatically analyze the behavior of the participants, through statistical fitting techniques, clustering, and time series analysis. These techniques will enable the automation of the evaluation tasks and the provision of detailed feedback, which would require a huge effort from the white teams.

\subsection{Actor simulation through AI techniques}
\label{sec:phase3}
The last phase of the research proposal will concern the development of software agents capable of simulating the actions of the actors, both from the offensive (connection attempts, traffic sniffing, vulnerability scanning) and defensive side (firewall, network segmentation, and services reconfiguration). These automated actors will be controlled through machine learning algorithms, such as \textit{reinforcement learning}, which will carry out specific sequences of actions to maximize the probability of bringing the system in the desired state. A solution based on such techniques is CyberBattleSim \cite{CyberBattleSim}, developed by Microsoft. This paradigm is similar to the one used by artificial intelligence systems to learn how to win a competition, e.g., AlphaGo by Google DeepMind \cite{DeepMind}. To increase the effectiveness of such techniques, the aforementioned monitoring system will be used to provide feedback to the algorithms to guide their training. 

Actor simulation is closely related to the concept of adversary emulation: a set of tactics, techniques, and procedures (TTPs) act to simulate the presence and the actions of a hostile subject inside the network \cite{MITREGettingStarted}. Adversary emulation is useful to measure the ability of the network itself to identify threats and prevent damages to its devices. Adversary emulation tools should be intelligent, realistic, modular and they should grant a low overhead. Adversary emulation represents an advantage for both red and blue teams. Like Cyber Range-based engagements, Adversary emulation activities are costly and time consuming, involving a significant amount of staff to set up the red team and identify the adversary's tactics, techniques and procedures (TTPs). To address this issue, many solutions known as Automated threat emulators have been proposed. Zilberman et al. \cite{zilberman2020sok} carried out an exhaustive analysis of the most popular threat emulators, ranked according to several criteria, e.g. coverage of the MITRE ATT\&CK Enterprise Matrix \cite{MITREMatrix}. Among these threat emulators we find MITRE CALDERA \cite{CALDERA}, \cite{applebaum2016intelligent}, Atomic Red Team \cite{AtomicRedTeam}, Red Team Automation \cite{RedTeamAutomation} and Metasploit \cite{Metasploit}. 

These solutions are very useful to automatically test the offensive and defensive strategies during training. However, there is room for further improvement of such tools. To address this matter, the study will focus on the integration of AI techniques to enable the system to learn how to defend the infrastructure from all types of threat agents.


\subsection{Status of the dissertation research}
The dissertation research is still at an early stage. During the first year, I will develop my own techniques to create virtualized environments for representative Cyber Ranges, along with the experimentation of the current adversary emulation tools.
During the second year, my work will focus on hypervisor-level techniques for monitoring and attack simulation. Finally, during the third year, I will work on the AI techniques for actor simulation.



%% file: tex/conclusion.tex
\section{Conclusion}
\label{sec:Conclusion}
In this paper, we describe a research proposal for the development of a next-generation Cyber Range platform. This novel platform aims to facilitate the creation and management of Cyber Ranges and to reduce the related costs. 
Our proposal consists of three different phases concerning a critical aspect of the platform: the advanced scenario simulation, the monitoring techniques, and, finally, the actor simulation. 
The platform will allow the companies to test the cyber-security aspects of their infrastructure on a virtual clone in order to provide high flexibility, ease of management and configuration, highly realistic scenario generation, automatic monitoring, and actor simulation. 

%% file: tex/acknowledgments.tex
\section*{Acknowledgment}
\label{sect:acks}
The disseration research is supervised by professors Domenico Cotroneo and Roberto Natella. This work has been partially supported by the Italian Ministry of University and Research (MUR) under the programme ``PON Ricerca e Innovazione 2014-2020 – Dottorati innovativi con caratterizzazione industriale''.

%% file: conference_101719.bbl
\begin{thebibliography}{10}

\bibitem{applebaum2016intelligent}
Andy Applebaum, Doug Miller, Blake Strom, Chris Korban, and Ross Wolf.
\newblock Intelligent, automated red team emulation.
\newblock In {\em Proceedings of the 32nd Annual Conference on Computer
  Security Applications}, pages 363--373, 2016.

\bibitem{carbone2014vprobes}
Martim Carbone, Alok Kataria, Radu Rugina, and Vivek Thampi.
\newblock Vprobes: deep observability into the esxi hypervisor.
\newblock {\em VMware Tech J}, 14(5):35--42, 2014.

\bibitem{cohen2005instrumenting}
W~Cohen.
\newblock Instrumenting the linux kernel with systemtap.
\newblock {\em Red Hat Magazine}, 2005.

\bibitem{cohen2005gaining}
William Cohen.
\newblock Gaining insight into the linux kernel with kprobes.
\newblock {\em RedHat Magazine}, (5), 2005.

\bibitem{VirginiaCR}
{Commonwealth of Virginia}.
\newblock {Virginia Cyber Range}.
\newblock \url{https://www.virginiacyberrange.org/}.

\bibitem{8390705}
D.~{Cotroneo}, L.~{De Simone}, and R.~{Natella}.
\newblock Run-time detection of protocol bugs in storage i/o device drivers.
\newblock {\em IEEE Transactions on Reliability}, 67(3):847--869, 2018.

\bibitem{DETER}
{Department of Homeland Security, Department of Justice}.
\newblock {The DETER Project}.
\newblock \url{https://deter-project.org/}.

\bibitem{RedTeamAutomation}
{Endgame Inc.}
\newblock {Red Team Automation}.
\newblock \url{https://github.com/endgameinc/RTA}.

\bibitem{garfinkel2003virtual}
Tal Garfinkel, Mendel Rosenblum, et~al.
\newblock A virtual machine introspection based architecture for intrusion
  detection.
\newblock In {\em Ndss}, volume~3, pages 191--206. Citeseer, 2003.

\bibitem{Redcloud}
{GitHub}.
\newblock {Redcloud}.
\newblock \url{https://github.com/khast3x/Redcloud}.

\bibitem{DeepMind}
{Google DeepMind}.
\newblock {AlphaGo}.
\newblock
  \url{https://deepmind.com/research/case-studies/alphago-the-story-so-far}.

\bibitem{gregg2011dtrace}
Brendan Gregg and Jim Mauro.
\newblock {\em DTrace: Dynamic Tracing in Oracle Solaris, Mac OS X, and
  FreeBSD}.
\newblock Prentice Hall Professional, 2011.

\bibitem{KVM}
KVM.
\newblock {KVM}.
\newblock
  \url{https://www.linux-kvm.org/index.php?title=Main_Page&oldid=173792}.

\bibitem{systems7010007}
Azad~M. Madni, Carla~C. Madni, and Scott~D. Lucero.
\newblock Leveraging digital twin technology in model-based systems
  engineering.
\newblock {\em Systems}, 7(1), 2019.

\bibitem{KYPO}
{Masaryk University}.
\newblock {KYPO}.
\newblock \url{https://crp.kypo.muni.cz/}.

\bibitem{MichiganCR}
{Merit Network}.
\newblock {Michigan Cyber Range}.
\newblock \url{https://www.merit.edu/security/training/hubs/}.

\bibitem{Metasploit}
{Metasploit}.
\newblock {Metasploit}.
\newblock \url{https://www.metasploit.com/}.

\bibitem{CyberBattleSim}
{Microsoft}.
\newblock {Gamifying machine learning for stronger security and AI models}.
\newblock
  \url{https://www.microsoft.com/security/blog/2021/04/08/gamifying-machine-learning-for-stronger-security-and-ai-models/}.

\bibitem{CALDERA}
{MITRE}.
\newblock {CALDERA}.
\newblock
  \url{https://www.mitre.org/research/technology-transfer/open-source-software/caldera\%E2\%84\%A2}.

\bibitem{MITREGettingStarted}
{MITRE}.
\newblock {Getting started with ATT\&CK}.
\newblock
  \url{https://www.mitre.org/sites/default/files/publications/mitre-getting-started-with-attack-october-2019.pdf}.

\bibitem{MITREMatrix}
{MITRE}.
\newblock {MITRE ATT\&CK Enterprise Matrix}.
\newblock \url{https://attack.mitre.org/matrices/enterprise/}.

\bibitem{NIST-CR-Guide}
{National Institute of Standards and Technology}.
\newblock {The Cyber Range: A Guide}.
\newblock
  \url{https://www.nist.gov/system/files/documents/2020/06/25/The\%20Cyber\%20Range\%20\-\%20A\%20Guide\%20\%28NIST-NICE\%29\%20\%28Draft\%29\%20-\%20062420\_1315.pdf},
  2020.

\bibitem{GENI}
{National Science Foundation}.
\newblock {GENI}.
\newblock \url{https://www.geni.net/}.

\bibitem{OpenStack}
{Open Infrastructure Foundation}.
\newblock {OpenStack}.
\newblock \url{https://www.openstack.org/}.

\bibitem{NCR}
{PEO STRI}.
\newblock {NCR}.
\newblock \url{https://www.peostri.army.mil/national-cyber-range-ncr}.

\bibitem{pfoh2009formal}
Jonas Pfoh, Christian Schneider, and Claudia Eckert.
\newblock A formal model for virtual machine introspection.
\newblock In {\em Proceedings of the 1st ACM workshop on Virtual machine
  security}, pages 1--10, 2009.

\bibitem{AtomicRedTeam}
{Red Canary}.
\newblock {Atomic Red Team}.
\newblock \url{https://atomicredteam.io/}.

\bibitem{staff2012joint}
US~Joint Staff.
\newblock Joint training manual for the armed forces of the united states
  (cjcsm 3500.03 d).
\newblock {\em Washington, DC: Joint Chiefs of Staff}, 2012.

\bibitem{8477101}
F.~{Tao}, H.~{Zhang}, A.~{Liu}, and A.~Y.~C. {Nee}.
\newblock Digital twin in industry: State-of-the-art.
\newblock {\em IEEE Transactions on Industrial Informatics}, 15(4):2405--2415,
  2019.

\bibitem{VMware-ESXi}
{VMware}.
\newblock {VMware ESXi}.
\newblock \url{https://www.vmware.com/it/products/esxi-and-esx.html}.

\bibitem{Detours}
Redmond Wa, Galen Hunt, and Doug Brubacher.
\newblock Detours: Binary interception of win32 functions.
\newblock {\em Proceedings of the 3rd Conference on USENIX Windows NT
  Symposium}, 3, 02 1970.

\bibitem{AYAMIN2020101636}
Muhammad~Mudassar Yamin, Basel Katt, and Vasileios Gkioulos.
\newblock Cyber ranges and security testbeds: Scenarios, functions, tools and
  architecture.
\newblock {\em Computers \& Security}, 88:101636, 2020.

\bibitem{zilberman2020sok}
Polina Zilberman, Rami Puzis, Sunders Bruskin, Shai Shwarz, and Yuval Elovici.
\newblock Sok: A survey of open-source threat emulators.
\newblock {\em arXiv preprint arXiv:2003.01518}, 2020.

\end{thebibliography}
